\documentclass[journal]{IEEEtran}
\usepackage[square,sort&compress,numbers]{natbib}

\usepackage{amssymb}
\usepackage[ruled,linesnumbered]{algorithm2e}
\usepackage{algorithmic}
\usepackage{amsmath,amsthm}
\usepackage{graphicx}
\usepackage{epsfig}
\usepackage{subfigure}
\usepackage{url}
\usepackage{multirow}
\usepackage{graphicx} %插入图片的宏包
\usepackage{float} %设置图片浮动位置的宏包
\usepackage{stfloats}

\usepackage{subfigure}
\usepackage{color}
\usepackage{xcolor}
\usepackage[numbers,sort&compress]{natbib}
\graphicspath{{pic/}}
\ifCLASSINFOpdf
\else
\fi

\begin{document}
\title{Routing in Small Satellite Networks: A GNN-based Learning Approach}

\author{\IEEEauthorblockN{Mengjie Liu, Jian Li,~\IEEEmembership{Member,~IEEE,}, Hancheng Lu,~\IEEEmembership{Senior Member,~IEEE}}

\IEEEcompsocitemizethanks{\IEEEcompsocthanksitem
M. Liu, J. Li and H. Lu are with the University of Science and Technology of China, Hefei 230027, China (Email: lmj316@mail.ustc.edu.cn; lijian9@ustc.edu.cn; hclu@ustc.edu.cn).
}
}

\maketitle

\begin{abstract}
Small satellite networks (SSNs), which are constructed by large number of small satellites in low earth orbits (LEO), are considered as promising ways to provide ubiquitous Internet access. To handle stochastic Internet traffic, on-board routing is necessary in SSNs. However, large-scale, high dynamic SSN topologies and limited resources make on-board routing in SSNs face great challenges. To address this issue, we turn to graph neural network (GNN), a deep learning network inherently designed for graph data, motivated by the fact that SSNs can be naturally modeled as graphs. By exploiting GNN's topology extraction capabilities, we propose a GNN-based learning routing approach (GLR) to achieve near-optimal on-board routing with low complexity. We design high-order and low-order feature extractor and cross process to deal with high dynamic topologies of SSNs, even those topologies that have never been seen in training. Simulation results demonstrate that GLR results in a significant reduction in routing computation cost while achieves near-optimal routing performance in SSNs with different scales compared with typical existing satellite routing algorithms.

\end{abstract}


\IEEEpeerreviewmaketitle

\section{Introduction}
Due to the low manufacturing and deployment costs, a large number of small satellites in low earth orbits (LEO) are planed or have been launched to provide low-latency, high-bandwidth communication coverage, especially for areas that are difficult to construct terrestrial networks \cite{radtke2017interactions}, such as underdeveloped suburbs, deserts, oceans, etc \cite{talgat2020stochastic,zhu2020stochastic}. To break capability and resource limitations of a single small satellite and provide ubiquitous Internet access, small satellites which are usually operated in the form of constellations are expected to constitute networks through inter-satellite links \cite{su2019broadband}.

Small satellite networks (SSNs) will be integrated into future Internet, which supply alternative data packet forwarding paths for users in terrestrial networks \cite{ji2020popularity,fu2020secure}. In detail, data packets from source ground users are transmitted to SSNs over ground-to-satellite links, forwarded in SSNs, and then sent back to destination ground users by satellite-to-ground links. For SSNs, the most important task is to establish paths for data packet forwarding from source small satellites to destination small satellites, also known as satellite routing.

There have been many research attempts on satellite routing \cite{werner1997dynamic,taleb2008explicit,huang2016optimized,liu2015low, li2019temporal}, including scheduled routing and opportunistic routing. However, existing routing schemes are inappropriate for SSNs.
The rapid movement of small satellites will increase the complexity of the network topology and status of inter-satellite links, leading to inaccurate network topology pre-calculated in scheduled routing schemes \cite{werner1997dynamic,huang2016optimized}. Therefore, on-board routing is necessary in SSNs. Opportunistic routing schemes \cite{taleb2008explicit} collect network state information on-board and then perform routing computing based on collected information. However, the enormous overhead makes this kind of routing schemes unable to be applied to SSNs. SSNs are large in scale in order to realize global coverage. For example, Oneweb and SpaceX have launched or plan to launch hundreds or even more small satellites. Large scale together with highly dynamic nature imposes tremendous resource demand on opportunistic routing in SSNs. On the other hand, resources in small satellites are limited to ensure low cost. Hence, in large-scale SSNs, there is a significant gap between resources required for on-board routing and resources that small satellites can provide.

As mentioned above, high efficient on-board routing is most desired in SSNs to combat challenges brought by large-scale, high dynamic topologies and limited resources. Fortunately, the rapid development of deep learning over recent years has provided potential solutions with low complexity. In order to make use of the benefits from the predictable and periodic satellite's trajectory, SSNs are usually described as graphs. However, traditional neural networks used in deep learning such as convolutional neural networks (CNN), recurrent neural networks (RNN) are not suitable to deal with graph-structured information. Consequently, they are unable to learn optimal routing when applied in SSNs.

Different from traditional neural networks, GNN \cite{wu2019simplifying} is inherently designed for processing graph data. In this paper, we propose a GNN-based learning approach for routing in SSNs, by exploiting the GNN's topology extraction capabilities. The processing of GNN under a single topology is expanded for SSNs. To obtain powerful feature capture and reasoning capabilities, we have carefully designed high-order and low-order feature extractor and cross process. The designed modules can deal with different topologies of SSNs during the routing process, even those topologies that have never been seen in training. By offline pre-learning, on-board decision-making complexity can be greatly reduced. Therefore, efficient on-board routing can be achieved with large-scale, high dynamic SSN topology. Furthermore, we carry out simulations to evaluate the performance of the proposed routing approach in SSNs with different scales. Results show that the proposed routing approach can significantly reduce the complexity of routing computation while achieve near-optimal routing performance compared with typical existing satellite routing algorithms.

The rest of the paper is organized as follows. In Section II, we explain the formulation of SSNs routing problem, and in Section III, we introduce the framework and implementation of the proposed GNN-based routing approach for SSNs in detail. To verify the performance and computational cost of GRL, simulation results are shown in Section IV. Finally, we conclude our work in Section V.

\section{Problem Statement}
In this section, we describe the routing problem in SSNs and emphasize the difference between the traditional training process of learning-based routing approach and the proposed one.

The satellite network topology can be expressed as $G$, and its adjacency matrix is recorded as $A\in\mathbb{N}^{N\times N}$, where $N$ represents the total number of satellite nodes. In specific, $G=\{V,E,f(e)\}$, where $V$ denotes the set of satellite nodes, $E$ denotes the set of inter-satellite links at the current moment, and $f(e)$ is the link feature of link $e\in E$. Note that we model the predictable inter node delay as link feature $f(e)$ so as to enable each node to perform routing calculation independently.

\begin{figure}[t]
\centering 
\subfigure[node selection probability]{
\label{labeled_data1}
\includegraphics[width=0.2\textwidth]{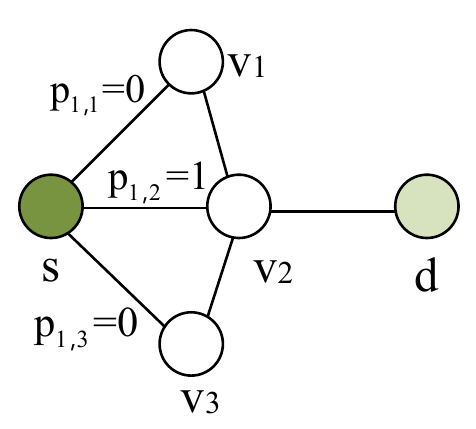}
}
\subfigure[communication distance]{
\label{labeled_data2}
\includegraphics[width=0.2\textwidth]{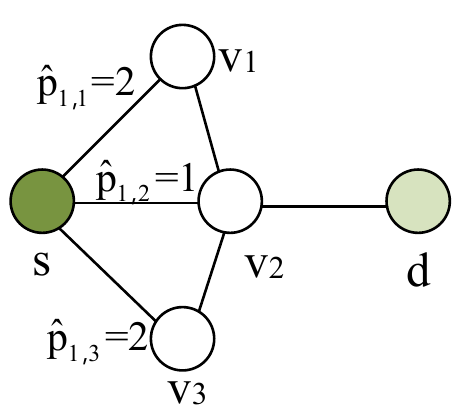}
}
\caption{Different label method used in training process, there are three neighbor nodes for next-hop selection at first-hop $t_1=s$ in this example.}
\label{labeled_data} 
\end{figure}

For a given source and destination pair $\langle s,d \rangle$, the routing process can be formulated as a series of next-hop path selection process, i.e., $R(s, d) = (t_1, t_2,...,t_n)$, where $R(s, d)$ denotes the routing path, and $t_i$ denotes the $i$-th hop in the routing path for $\langle s,d \rangle$.
Thus, the routing decision on the next hop $t_{i+1}$ should be performed iteratively, and the decision of $i+1$-th hop can be described as $t_{i+1}=P (t_i, d, G) $, which means the current hop $t_i$, destination $d$, and network topology information are considered during the selection process.

Different from the classical routing algorithms, e.g., greedy searching in Dijkstra algorithm, GLR is a data-driven approach, which requires labeled data and training process. Most of existing works formulate routing decision as the node selection probability of the next hop\cite{zhuang2019toward}. For $i$-th hop $t_i$, next-hop decision $P(t_i, d, G) = \arg\max(p_{i,1}, p_{i,2},...,p_{i,M})$, where $p_{i,m}$ denotes the probability of $v_m$, and $M$ denotes the number of inter-satellite links connected to $t_i$. As shown in Fig. \ref{labeled_data1}, $v_2$ points to the target next hop and $R(s, d) = (s,v_2,d)$ is the shortest path, then $p_{i,2}=1$, and $p_{i,1}=p_{i,3}=0$. However, by adopting such processing method, the result of the data set is always in the form of one hot encoding vector, which cannot provide sufficient information. Furthermore, when multiple “optimal paths” exist, e.g., multiple paths share the same delay or distance to the destination, the probability-based method becomes difficult to model and be used for training process.

Thus, rather than node selection probability, we consider the communication distance, which is the hop count between the current and destination node, as the output of our GLR model. As shown in Fig. \ref{labeled_data2}, next-hop decision $P(t_1, d, G)=\arg\min(\hat{p}_{1,1}, \hat{p}_{1,2}, \hat{p}_{1,3})=v_2$, where $t_1=s$. In this case, not only the information of the optimal node selection is recorded ($\hat{p}_{1,2}=1$), but also the information of the other node selections can be captured during the training process ($\hat{p}_{1,1}=\hat{p}_{1,3}=2$). By replacing node selection probability with communication distance, the classification problem in routing process can be converted into a regression problem to capture more feature information and obtain better routing performance.

\section{GNN-based Routing Approach}
\begin{figure*}[t]
\centering 
\includegraphics[width=0.9\textwidth]{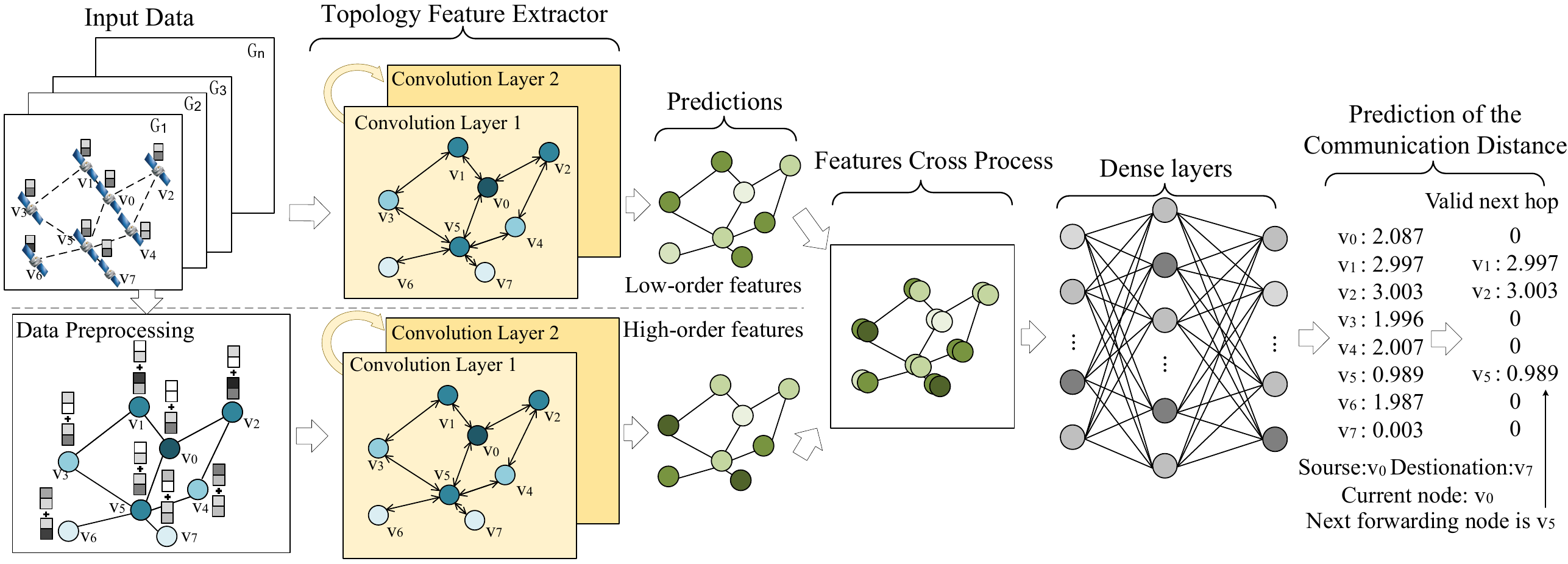} 
\caption{The overall structure of the proposed routing model. The sampled satellite data is preprocessed, and passes through multiple graph convolution layers. Then high-order and low-order topological features are crossed. At last, processed data passes through the traditional dense structures to obtain prediction results.}
\label{Fig.main1}
\end{figure*}

In this section, we describe the specific design of our GLR model in the following three aspects: 1)we design a topological feature extractor using the information aggregation ability of Graph Convolution Network (GCN). 2) we define high-order and low-order topological features, where high-order topological features are extracted by combining the original input with primary topological information. 3) we cross the above high-order and low-order topological features to obtain richer global information and then predict the node communication distance through the dense layer. The structure diagram of the proposed routing model is shown in Fig. \ref{Fig.main1}.

\subsection{Topological Feature Extractor}

Combining the input data and preprocessed data, the graph convolution mechanism is adopted to spread information.
Similar to CNN, GCN can spread farther information between nodes through multi-layers.

As shown in the middle part of Fig. \ref{Fig.main1}, the topological feature extractor is composed of two convolution layers.
In each convolution layer, the node information aggregation is performed by considering the adjacency relation and the node degree (i.e., the number of connected neighbors).
In specific, the information aggregation process of each node can be calculated by:
\begin{equation}
\overline{h}_i^l\leftarrow \frac{1}{\tilde{d}_i+1}h_i^{l-1}+\sum_{j=0}^{{N-1}}\frac{a_{ij}}{\sqrt{(\tilde{d}_i+1)(\tilde{d}_j+1)}}h_j^{l-1},  \label{GCN}
\end{equation}
where $h^i_l$ denotes the $i$-th node's output on the $l$-th layer, $a_{ij}\in \{0,1\}$ denotes the connection relation betweens node $v_i$ and node $v_j$, $\tilde{d}_i$ denotes the degree of node $v_i$.
In particular, the weight of a single node should be smaller when the node degree is larger.

Multiplied by the weighting matrix W and then updated by the relu function, the result can be finally written into matrix form, as is shown in eq. \eqref{GCN_f}.
\begin{equation}
\overline{H}^l\leftarrow \sigma{(\widetilde{D}^{-\frac{1}{2}} \widetilde{A}\widetilde{D}^{-\frac{1}{2}}H^{l-1}W^l )}, \label{GCN_f}
\end{equation}
where $\overline{H}^l$ denotes the output of $l$-th layer,
$\widetilde{A}=A+I$ is the adjacency matrix that adds self-connection which is corresponding to the first part in eq. \eqref{GCN}, $\widetilde{D}$ represents the degree diagonal matrix of $\widetilde{A}$, and $\sigma(\cdot)$ denotes relu function. 
Note that all nodes share the same $W$ to capture common features during the training process.

\subsection{High-Order and Low-Order Topological Features}
Due to the highly dynamic characteristic of the network topology in SSNs, the connection relations between satellite nodes are varying rapidly at different intervals, and there are few continuity characteristics in the time domain. To meet the routing requirement under such dynamic environment, stronger generalization ability is required in our routing model.

To capture more topological information from labeled data, a common solution is to adopt deeper graph convolutional network. However,
too deep layers will lead to excessively smooth information and poor results.
Generally speaking, 3 to 4 layers are the limit of the depth of GNN-based model.
Compared with CNN, this kind of depth is still difficult to capture the high-level structural information of the graph structure.
To tackle such problem, we designed two modes, i.e, low-order mode and high-order mode respectively, based on the topological feature extractor. The former one is directly obtained from original input through feature extractor, while the former one is obtained by another feature extractor after splicing preprocessed structure information. By doing so, the and the generalization ability of our model can be improved.

The specific splicing process in data preprocessing can be described as follows: Since the farther neighbor contains less effective information, 1 and 0.5 are used to represent the one-hop and two-hop neighbor of the node, respectively. After that, the output of data preprocessing becomes the input of the topological feature extractor in high-order mode.

\subsection{Features Cross and Prediction Process}
To take advantage of high-order and low-order topological features, we make features cross between them to obtain richer and wider network information.
Generally speaking, there are three common ways to achieve features cross, i.e., splicing, Hadaman product and outer product. In specific, splicing is the most basic feature intersection method, while Hadaman product and outer product can achieve node-level and dimension-level feature intersection with higher computational and parameter cost, respectively. Since the efficiency of routing model is an important aspect to resource-limited SSNs, splicing method, as the most efficient features cross method, is adopted in features cross process.

After features cross process, the data passes through the dense layer and  the final predicted result can be obtained. We use the communication distance as the predicted label as shown in the output part of Fig. \ref{Fig.main1}. Compared with the traditional next-hop prediction, the communication distance prediction is a regression problem, which contains multi-level, richer label information, and stronger guiding significance for model training.
The regression error $L$ is defined by a loss function:
\begin{equation}
L =  \frac{1}{n}\sum_{i=1}^{n}(y_i - y_i^p)^2 + \beta||\theta||_2,   \label{loss}
\end{equation}
where $y_i$ and $y_i^p$ represent the predicted communication distance and ground-truth distance, respectively. In order to avoid over-fitting, a regularization term $\theta$ is also added, where $\beta$ represent the significance of the regularization part.

\begin{table}[h]
\renewcommand{\arraystretch}{1.1}
\caption{Simulation Parameters}
	\centering
	\begin{tabular}{p{6.5cm}l}
	%\centering
	\hline
	\textbf{Parameter}&\textbf{Value}\\
	\hline
    LEO orbit height&1050km\\
	%LEO orbit height&550km-1050km\\
	Number of LEO satellites&132-396\\
	Number of orbital planes&12-36\\
	Eccentricity&0\\
	Inclination&$53^\circ$\\
	Maximum elevation angle&$50.4^\circ$\\
	%Buffer size&16Mb\\
	Transmission rate&100kps\\
%	Total generated packets&5000\\
	Communication range&3500km\\
	\hline
	\end{tabular}
\label{table}
\end{table}
\section{Performance Evaluation}
\subsection{Simulation Setup}
In this section, a serious of simulations are conducted to evaluate the performance of the proposed GLR approach, and simulation parameter settings are summarized in the Table \ref{table}.
To construct an Iridium-like constellation, an open-sourced simulation platform called LSSN \cite{li2019temporal,9350838} is adopted.
Hardware configurations of our simulation platform are: Core i7-7700 CPU, 3.6 Ghz, 8G RAM, O.S. Windows 10 Professional 64 bits. Similar to \cite{li2019temporal}, the following routing algorithms are adopted as comparisons:

\begin{itemize}
\item
\textbf{Temporal graph based Brute-force Routing (TBR):} TBR adopts brute-force algorithm at each hop to find the optimal path.
\item
\textbf{Temporal graph based Source Routing (TSR):} TSR executes Dijkstra algorithm once at the source node and encapsulates the calculated path in the packet header. The data packet will be forwarded according to the pre-calculated path during the transmission.
\item
\textbf{Contact Graph Routing (CGR):} CGR utilizes Dijkstra algorithm to find the shortest routing path from pre-planned links (i.e., scheduled contact plan).
\end{itemize}
\begin{figure}[t]
  \centering
\subfigure[performance in terms of average delay]{
\label{performance1}
\includegraphics[width=0.3\textwidth]{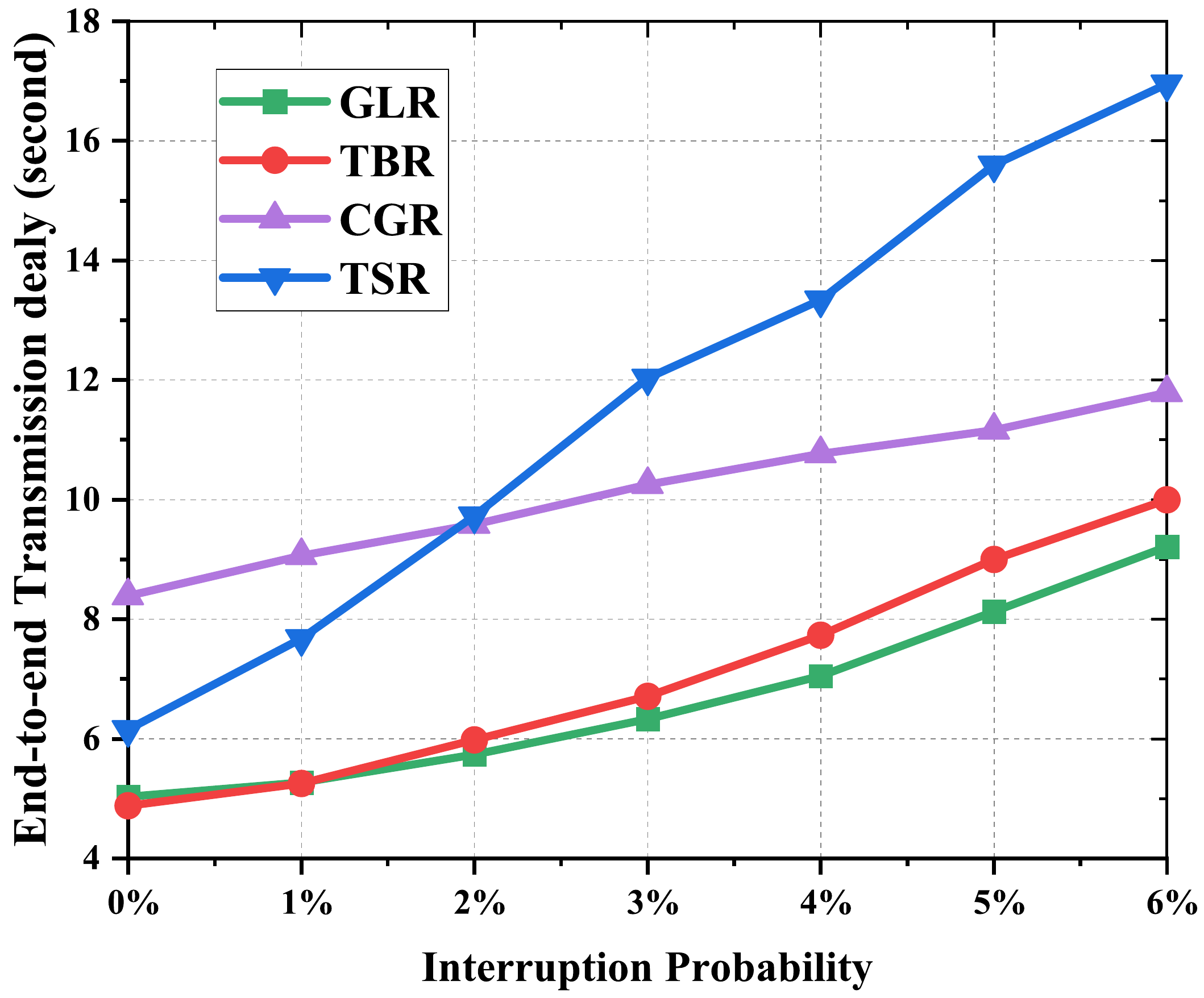}
}
\subfigure[performance in terms of packet drop rate]{
\label{performance2}
\includegraphics[width=0.305\textwidth]{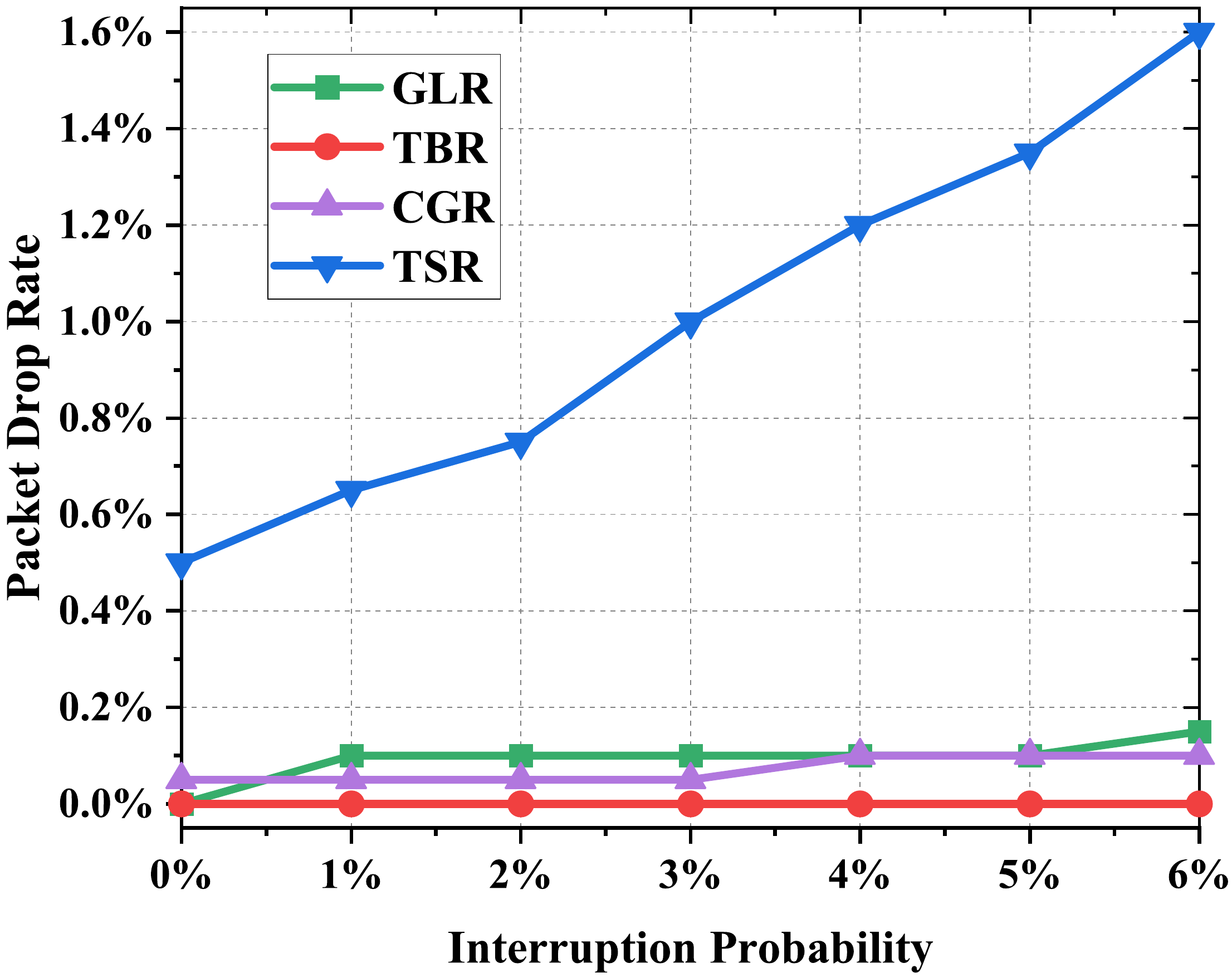}
}
\caption{Performance comparison vs. interruption probability.}
\label{Fig.3}
\end{figure}

\subsection{Routing Performance}

Considering the complex communication environment of SSNs, we evaluate the performance of the proposed GLR approach and the traditional routing algorithms under different link interruption probabilities as shown in Fig. \ref{Fig.3}. Since the higher link interruption leads to the increases of rerouting probabilities, the performance of all routing approaches in terms of delay and packet drop rate is decreasing with the increase of interruption probability. Compared with TSR, other routing algorithms can obtain alternative routing solutions through rerouting, so they can relatively get better performance. In particular, facing link interruptions, GLR shows stronger robustness with its topology information capture mechanism.

\begin{figure}[t]
\centering 
	\includegraphics[width=0.33\textwidth]{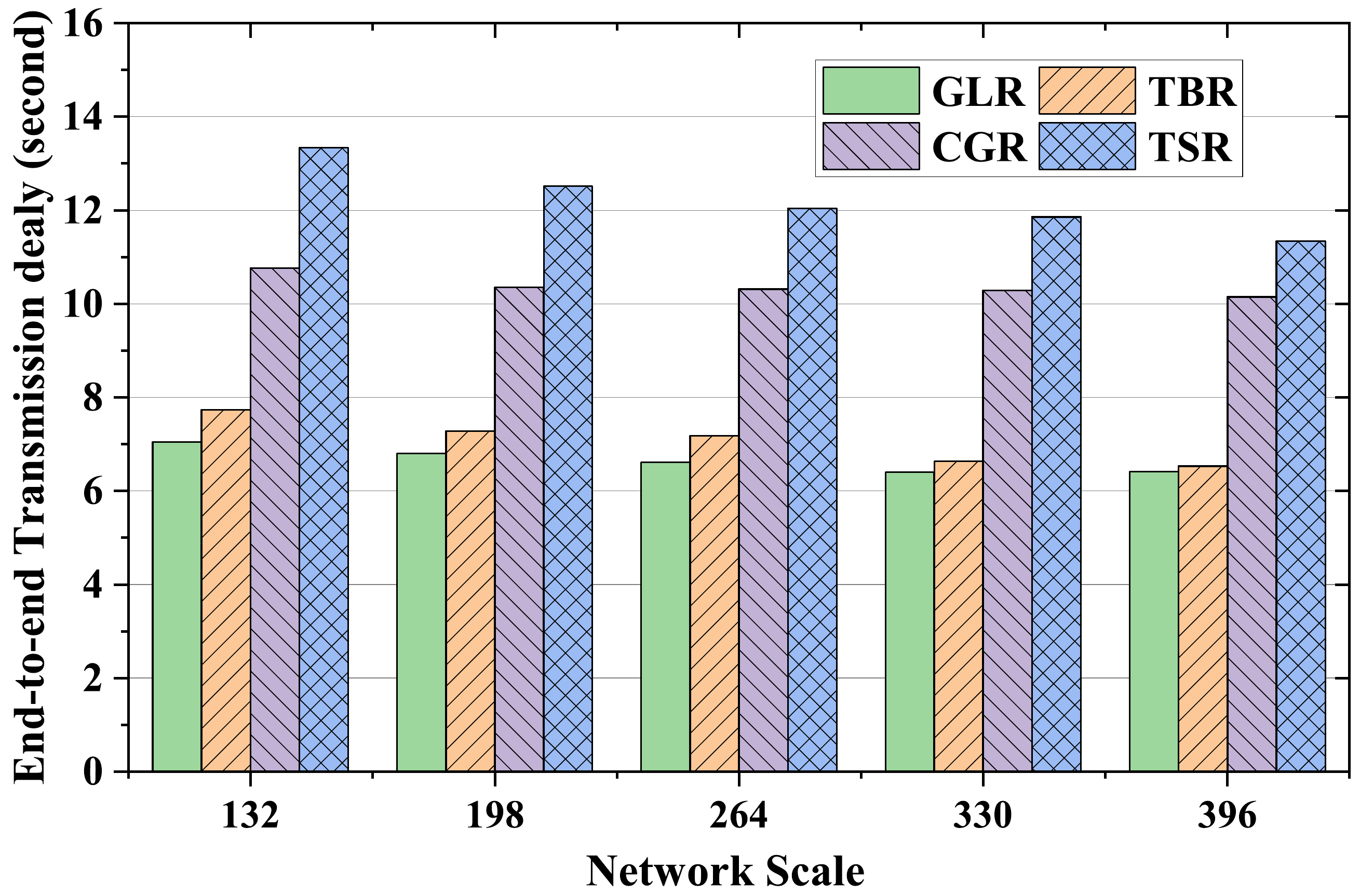} 
	\caption{Transmission delay vs. network scale (Interruption probability = 4\%.)} 
\label{performance3} 
\end{figure}

\begin{figure}[t]
\centering 
	\includegraphics[width=0.33\textwidth]{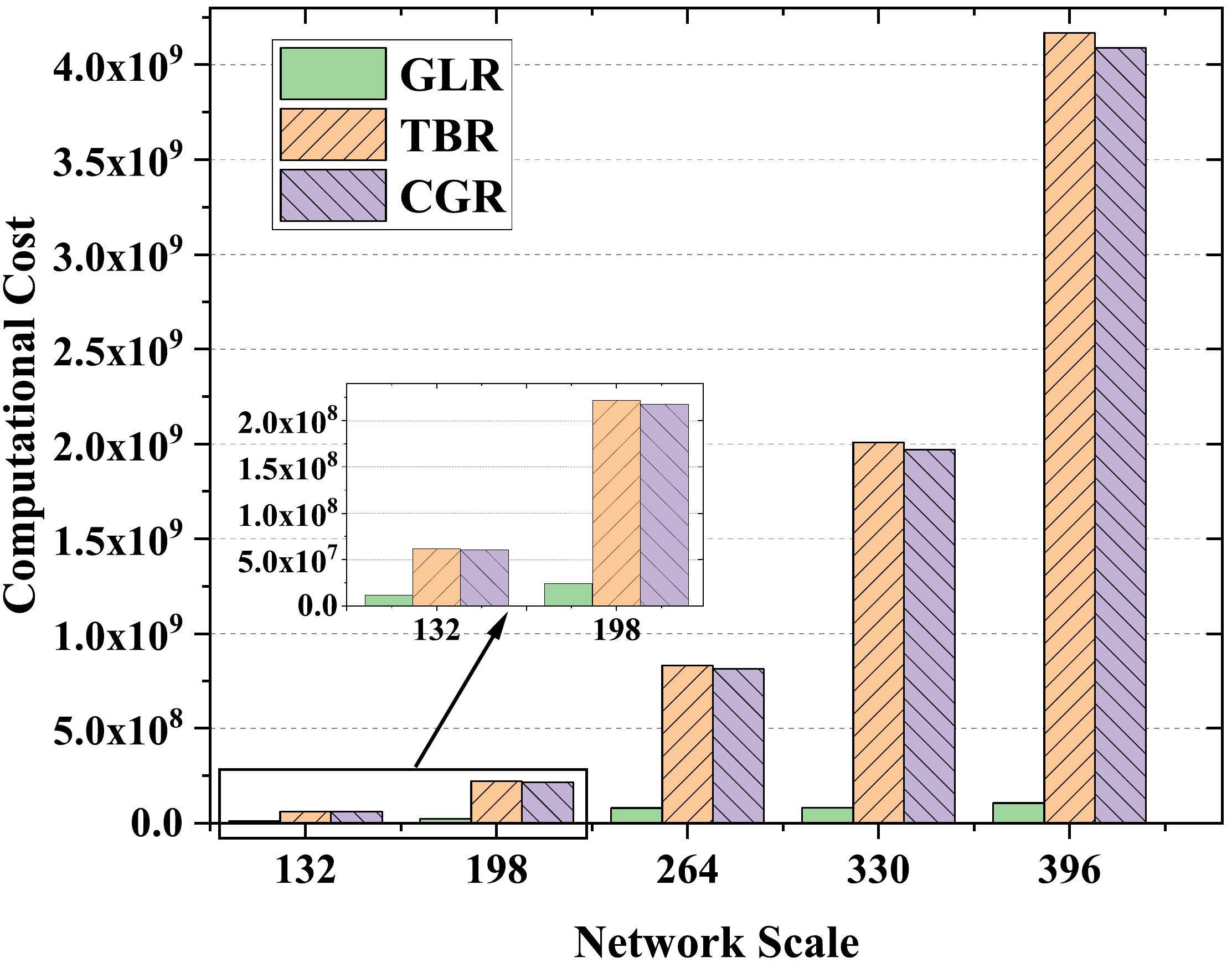} 
	\caption{Computational cost vs.  network scale.} 
\label{Computation}
\end{figure}

As shown in Fig. \ref{performance3}, with the increase of network scale, the proposed GLR approach achieves better performance and the end-to-end transmission delay is decreasing. This phenomenon implies that GLR can adapt to various network topology.
In Fig. \ref{Computation}, we further compare the computational cost of each routing algorithm.
By measuring the execution time of each algorithm, we observe that the computational cost of TBR and CGR rises drastically with the increase of network scale, while GLR grows slowly at a lower level in Fig. \ref{Computation}. This phenomenon implies that GLR can reduce significantly computational overhead, which makes GLR become a better choice for on-board routing decisions in SSNs.

\section{Conclusion}

In this paper, we proposed a low-complexity on-board routing approach in SSNs. To address with the large-scale, high dynamic SSN topologies and limited resources at satellites, we defined the communication distance, thereby converting the classification problem into a regression one. By defining high-order and low-order feature extractor and feature cross process, the proposed GLR model can handle dynamic topologies or random interruptions due to its strong generalization ability. Simulation results validate that, GLR can obtain near-optimal routing performance with lower computational cost compared with traditional routing algorithms.
Integrating LEO networks and terrestrial networks, terrestrial-satellite networks face severe challenges in routing problem when facing frequent satellite-ground link switching. In the future, we can further explore the GNN-based routing approach in the terrestrial-satellite networks using GNN's topology processing capability.

\footnotesize
\bibliographystyle{ieeetr}
\bibliography{bibliography}

\begin{thebibliography}{10}

\bibitem{radtke2017interactions}
J.~Radtke, C.~Kebschull, and E.~Stoll, ``Interactions of the space debris
  environment with mega constellations—using the example of the oneweb
  constellation,'' {\em Acta Astronautica}, vol.~131, pp.~55--68, 2017.

\bibitem{talgat2020stochastic}
A.~Talgat, M.~A. Kishk, and M.-S. Alouini, ``Stochastic geometry-based analysis
  of leo satellite communication systems,'' {\em IEEE Communications Letters},
  2020.

\bibitem{zhu2020stochastic}
Y.~Zhu, D.~Zhou, M.~Sheng, J.~Li, and Z.~Han, ``Stochastic delay analysis for
  satellite data relay networks with heterogeneous traffic and transmission
  links,'' {\em IEEE Transactions on Wireless Communications}, vol.~20, no.~1,
  pp.~156--170, 2020.

\bibitem{su2019broadband}
Y.~Su, Y.~Liu, Y.~Zhou, J.~Yuan, H.~Cao, and J.~Shi, ``Broadband leo satellite
  communications: Architectures and key technologies,'' {\em IEEE Wireless
  Communications}, vol.~26, no.~2, pp.~55--61, 2019.

\bibitem{ji2020popularity}
Z.~Ji, S.~Wu, C.~Jiang, and W.~Wang, ``Popularity-driven content placement and
  multi-hop delivery for terrestrial-satellite networks,'' {\em IEEE
  Communications Letters}, vol.~24, no.~11, pp.~2574--2578, 2020.

\bibitem{fu2020secure}
Y.~Fu, C.-A. Jiang, Y.~Qin, and L.~Yin, ``Secure routing and transmission
  scheme for space-ocean broadband wireless network,'' {\em Science China
  Information Sciences}, vol.~63, no.~4, pp.~1--3, 2020.

\bibitem{werner1997dynamic}
M.~Werner, ``A dynamic routing concept for atm-based satellite personal
  communication networks,'' {\em IEEE journal on selected areas in
  communications}, vol.~15, no.~8, pp.~1636--1648, 1997.

\bibitem{taleb2008explicit}
T.~Taleb, D.~Mashimo, A.~Jamalipour, N.~Kato, and Y.~Nemoto, ``Explicit load
  balancing technique for ngeo satellite ip networks with on-board processing
  capabilities,'' {\em IEEE/ACM transactions on Networking}, vol.~17, no.~1,
  pp.~281--293, 2008.

\bibitem{huang2016optimized}
J.~Huang, Y.~Su, L.~Huang, W.~Liu, and F.~Wang, ``An optimized snapshot
  division strategy for satellite network in gnss,'' {\em IEEE Communications
  Letters}, vol.~20, no.~12, pp.~2406--2409, 2016.

\bibitem{liu2015low}
X.~Liu, X.~Yan, Z.~Jiang, C.~Li, and Y.~Yang, ``A low-complexity routing
  algorithm based on load balancing for leo satellite networks,'' in {\em 2015
  IEEE 82nd Vehicular Technology Conference (VTC2015-Fall)}, pp.~1--5, IEEE,
  2015.

\bibitem{li2019temporal}
J.~Li, H.~Lu, K.~Xue, and Y.~Zhang, ``Temporal netgrid model-based dynamic
  routing in large-scale small satellite networks,'' {\em IEEE Transactions on
  Vehicular Technology}, vol.~68, no.~6, pp.~6009--6021, 2019.

\bibitem{wu2019simplifying}
F.~Wu, A.~Souza, T.~Zhang, C.~Fifty, T.~Yu, and K.~Weinberger, ``Simplifying
  graph convolutional networks,'' in {\em International conference on machine
  learning}, pp.~6861--6871, PMLR, 2019.

\bibitem{zhuang2019toward}
Z.~Zhuang, J.~Wang, Q.~Qi, H.~Sun, and J.~Liao, ``Toward greater intelligence
  in route planning: A graph-aware deep learning approach,'' {\em IEEE Systems
  Journal}, vol.~14, no.~2, pp.~1658--1669, 2019.

\bibitem{9350838}
M.~Liu, Y.~Gui, J.~Li, and H.~Lu, ``Large-scale small satellite network
  simulator: Design and evaluation,'' in {\em 2020 3rd International Conference
  on Hot Information-Centric Networking (HotICN)}, pp.~194--199, 2020.

\end{thebibliography}

%\section*{Biographies}

\end{document}